\documentclass[reqno]{article}
\usepackage{amsmath,amssymb}
\newcommand\rf[1]{(\ref{eq:#1})}
\newcommand\lab[1]{\label{eq:#1}}
\newcommand\nonu{\nonumber}
\newcommand\br{\begin{eqnarray}}
\newcommand\er{\end{eqnarray}}
\newcommand\be{\begin{equation}}
\newcommand\ee{\end{equation}}

\newcommand\foot[1]{\footnotemark\footnotetext{#1}}
\newcommand\lb{\lbrack}
\newcommand\rb{\rbrack}

\renewcommand\({\left(}
\renewcommand\){\right)}
\renewcommand\v{\vert}                     
\newcommand\Bgv{\;\Bigg\vert}            
\newcommand\lskip{\vskip\baselineskip\vskip-\parskip\noindent}

\newcommand\bc{\begin{center}}
\newcommand\ec{\end{center}}

\newcommand\Tr{\mathop{\mathrm Tr}}                  
\newcommand\partder[2]{\frac{{\partial {#1}}}{{\partial {#2}}}}

\newcommand\Sbr[2]{\Bigl\lbrack\,{#1}\, ,\,{#2}\,\Bigr\rbrack}

\newcommand\Pbr[2]{\Bigl\{ \,{#1}\, ,\,{#2}\,\Bigr\}}  

\renewcommand\a{\alpha}

\renewcommand\d{\delta}

\newcommand\vareps{\varepsilon}
\newcommand\g{\gamma}
\newcommand\G{\Gamma}

\newcommand\h{\frac{1}{2}}
\renewcommand\k{\kappa}
\renewcommand\l{\lambda}
\renewcommand\L{\Lambda}
\newcommand\m{\mu}
\newcommand\n{\nu}
\renewcommand\o{\over}
\newcommand\om{\omega}
\renewcommand\O{\Omega}

\newcommand\vp{\varphi}
\renewcommand\P{\Phi}
\newcommand\pa{\partial}

\newcommand\pr{\prime}

\newcommand\s{\sigma}

\renewcommand\t{\tau}
\renewcommand\th{\theta}

\newcommand\wti{\widetilde}

\newcommand\cB{{\mathcal B}}
\newcommand\cC{{\mathcal C}}

\newcommand\cE{{\mathcal E}}

\newcommand\cL{{\mathcal L}}
\newcommand\cM{{\mathcal M}}
\newcommand\cN{{\mathcal N}}

\newcommand\cP{{\mathcal P}}

\newcommand\cT{{\mathcal T}}

\newcommand{\ct}[1]{\cite{#1}}
\newcommand{\bi}[1]{\bibitem{#1}}
\newcommand\PRD[3]{(#2), \textsl{Phys. Rev.} \textbf{D#1} #3}

\newcommand\PLB[3]{(#2), \textsl{Phys. Lett.} \textbf{#1B} #3}
\newcommand\CQG[3]{(#2), \textsl{Class. Quantum Grav.} \textbf{#1} #3}

\begin{document}

\title{String and Brane Tensions as Dynamical Degrees of Freedom}

\bigskip
\author{Eduardo Guendelman and Alexander Kaganovich\\
\small\it Department of Physics,\\[-1.mm]
\small\it Ben-Gurion University, Beer-Sheva, Israel  \\[-1.mm]
\small\it email: guendel@bgumail.bgu.ac.il, alexk@bgumail.bgu.ac.il \\
${}$ \\
Emil Nissimov and Svetlana Pacheva\\
\small\it Institute for Nuclear Research and Nuclear Energy,\\[-1.mm]
\small\it Bulgarian Academy of Sciences, Sofia, Bulgaria  \\[-1.mm]
\small\it email: nissimov@inrne.bas.bg, svetlana@inrne.bas.bg}
\date{ }


\maketitle
\bigskip

\begin{abstract}
We discuss a new class of string and $p$-brane models where the string/brane
tension appears as an {\em additional dynamical degree of freedom} instead of being
introduced by hand as an {\em ad hoc} dimensionfull scale. The latter
property turns out to have a significant impact on the string/brane dynamics.
The dynamical tension obeys Maxwell (or Yang-Mills) equations of motion 
(in the string case) or their rank $p$ gauge theory analogues (in the
$p$-brane case), which in particular triggers a simple classical mechanism
of (``color'') charge confinement.

\end{abstract}

\section{Introduction}

In order to build actions describing dynamics in geometrically motivated
field theories (for reviews of string and brane theories, see
\ct{string-brane-rev}) we need among other things a consistent
generally-covariant integration measure density, \textsl{i.e.}, covariant
under arbitrary diffeomorphisms (reparametrizations) on the underlying
space-time manifold. Usually the natural choice is the standard
Riemannian metric density $\sqrt{-g}$ with $g \equiv \det\v\v g_{\m\n}\v\v$.
However, there are no purely geometric reasons which prevent us from
employing an alternative generally-covariant integration measure. For instance,
introducing additional $D$ scalar fields $\vp^i$ ($i=1,\ldots ,D$ where $D$
is the space-time dimension) we may take the following new non-Riemannian
measure density $\P (\vp)$ :
\be
\P (\vp) \equiv {1\o {D!}}\vareps^{\m_1 \ldots \m_D}
\vareps_{i_1 \ldots i_D} \pa_{\m_1} \vp^{i_1} \ldots \pa_{\m_D} \vp^{i_D} \; .
\lab{D-measure-def}
\ee

Using \rf{D-measure-def} allows to construct new classes of models involving 
Gravity called \textsl{Two-Measure Gravitational Models} \ct{TMT}, whose
actions are typically of the form:
\be
S = \int d^D x\, \P (\vp)\, L_1 + \int d^D x\,\sqrt{-g}\, L_2 \; ,
\lab{TMT-a}
\ee
\be
L_{1,2} = e^{\frac{\a \phi}{M_P}} \Bigl\lb - {1\o \k} R(g,\G ) -
\h g^{\m\n}\pa_\m \phi \pa_\n \phi
+ \bigl(\mathrm{Higgs}\bigr) + \bigl(\mathrm{fermions}\bigr)\Bigr\rb \; .
\lab{TMT-b}
\ee
where $R(g,\G )$ is the scalar curvature in the first-order formalism
(\textsl{i.e.}, the affine connection $\G$ is independent of the metric),
$\phi$ is the dilaton field, $M_P$ is the Planck mass, \textsl{etc.}. 
Although naively the additional
``measure-density'' scalars $\vp^i$ appear in \rf{TMT-a} as pure-gauge
degrees of freedom (due to the invariance under arbitrary diffeomorphisms in
the $\vp^i$-target space), there is a remnant -- the so called ``geometric'' field
$\zeta (x) \equiv \frac{\P (\vp)}{\sqrt{-g}}$, which remains an additional
dynamical degree of freedom beyond the standard physical degrees of freedom
characteristic to the ordinary gravity models with the standard
Riemannian-metric integration measure. The most important property of the 
``geometric'' field $\zeta (x)$ is that its dynamics is determined only through the
matter fields locally (\textsl{i.e.}, without gravitational interaction).
The latter turns out to have a significant impact on the physical properties
of the two-measure gravity models which allows them to address various basic
problems in cosmology and particle physics phenomenology and provide
physically plausible solutions, for instance: (i) the issue of scale invariance
and its dynamical breakdown, \textsl{i.e.}, spontaneous generation of
dimensionfull fundamental scales; (ii) cosmological constant problem;  
(iii) geometric origin of fermionic families.  

For a recent review of two-measure gravity models see the contribution in this
volume \ct{TMT-Kiten}. In what follows we are going to apply the above ideas to
the case of string and $p$-brane models. Part of our exposition is based on
earlier works \ct{mstring}. Furthermore, we will elaborate on various important
properties of the modified-measure string and brane models with dynamical
string/brane tension.

\section{Bosonic Strings with a Modified World-Sheet Integration Measure}
We begin by first recalling the standard Polyakov-type action for the
bosonic string \ct{Polyakov}:
\be
S_{\mathrm{Pol}} = - T \int d^2\s\,
\h\sqrt{-\g}\g^{ab}\pa_a X^\m \pa_b X^\n G_{\m\n}(X) \; .
\lab{string-action-Pol}
\ee
Here $(\s^0,\s^1) \equiv (\t,\s)$; $a,b =0,1$; $\m,\n = 0,1,\ldots ,D-1$;
$G_{\m\n}$ denotes the Riemannian metric on the embedding space-time;
$\g_{ab}$ is the intrinsic Riemannian metric on the $1+1$-dimensional string 
world-sheet and $\g = \det\v\v\g_{ab}\v\v$; $T$ indicates the string tension 
-- a dimensionfull scale introduced \textsl{ad hoc}.
The resulting equations of motion w.r.t. $\g^{ab}$ and $X^\m$ read,
respectively:
\br
T_{ab}\equiv \(\pa_a X^\m \pa_b X^\n - 
\h \g_{ab} \g^{cd} \pa_c X^\m \pa_d X^\n\) G_{\m\n}(X) = 0 \; ,
\lab{gamma-eqs-Pol} \\
{1\o {\sqrt{-\g}}}\pa_a \(\sqrt{-\g}\g^{ab}\pa_b X^\m\) +
\g^{ab}\pa_a X^\n \pa_b X^\l \G^\m_{\n\l} = 0 \; ,
\lab{X-eqs-Pol}
\er
where $\G^\m_{\n\l}=\h G^{\m\k}\(\pa_\n G_{\k\l}+\pa_\l G_{\k\n}-\pa_\k G_{\n\l}\)$ 
is the affine connection for the external metric.

Let us now introduce two additional world-sheet scalar fields $\vp^i$ ($i=1,2$)
and replace $\sqrt{-\g}$ with a new reparametrization-covariant
world-sheet integration measure density $\P(\vp)$ defined in terms of $\vp^i$ :
\be
\P (\vp) \equiv \h \vareps_{ij} \vareps^{ab} \pa_a \vp^i \pa_b \vp^j
= \vareps_{ij} {\dot{\vp}}^i \pa_\s \vp^j \; .
\lab{def-measure}
\ee
However, the naively generalized string action
$S_{1} = -\h \int d^2\s\,\P (\vp) \g^{ab} \pa_a X^\m \pa_b X^\n G_{\m\n}(X)$
has a problem: the equations of motion w.r.t. $\g^{ab}$ lead to an unacceptable
condition $\P (\vp)\,\pa_a X^\m \pa_b X^\n G_{\m\n}(X) = 0$, \textsl{i.e.},
vanishing of the induced metric on the world-sheet.

To remedy the above situation let us consider topological (total-derivative) terms
w.r.t. standard Riemannian world-sheet integration measure. Upon measure replacement
$\sqrt{-\g} \to \P (\vp)$ the former are {\em not any more} topological --
they will contribute nontrivially to the equations of motion. For instance:
\be
\int d^2\s\,\sqrt{-\g}\, R  \to \int d^2\s\,\P (\vp) \, R \quad ,\;\;
R = \frac{\vareps^{ab}}{2\sqrt{-\g}} \(\pa_a \om_b - \pa_b \om_a\) \; ,
\lab{Euler-mod}
\ee
where $R$ is the scalar curvature w.r.t. $D=2$ spin-connection
$\om_{a}^{\bar{a}\bar{b}}=\om_{a}\vareps^{\bar{a}\bar{b}}$ (here $\bar{a},\bar{b}$
denote tangent space indices). The vector field $\om_{a}$ behaves as world-sheet
Abelian gauge field.

Eq.\rf{Euler-mod} prompts us to construct the following consistent modified
bosonic string action\foot{In ref.\ct{K-V} another interesting geometric
modification of the standard bosonic string model has been proposed, which
is based on dynamical world-sheet metric and torsion.} :
\be
S = - \int d^2\s \,\P (\vp) \Bigl\lb \h \g^{ab} \pa_a X^{\m} \pa_b X^{\n} G_{\m\n}
- \frac{\vareps^{ab}}{2\sqrt{-\g}} F_{ab}(A)\Bigr\rb \; ,
\lab{string-action}
\ee
where $\P (\vp)$ is given by \rf{def-measure} and 
$F_{ab}(A) \equiv \pa_{a} A_{b} - \pa_{b} A_{a}$ is the field-strength of an
auxiliary Abelian gauge field $A_a$. The action \rf{string-action} is
reparametrization-invariant as its ordinary string analogue \rf{string-action-Pol}. 
Furthermore, \rf{string-action} is invariant under diffeomorphisms in 
$\vp$-target space supplemented with a special conformal transformation of 
$\g_{ab}$ : 
\be
\vp^{i} \longrightarrow \vp^{\pr\, i} = \vp^{\pr\, i} (\vp) \quad ,\quad
\g_{ab} \longrightarrow \g^{\pr}_{ab} = J \g_{ab} \;\; ,\;\;
J \equiv \det \Bigl\Vert \frac{\pa\vp^{\pr\, i}}{\pa\vp^j} \Bigr\Vert \; .
\lab{Phi-Weyl-symm}
\ee
The latter symmetry, which we will call {\em ``$\P$-extended Weyl symmetry''},
is the counterpart of the ordinary Weyl conformal symmetry of the standard
string action \rf{string-action-Pol}.

The equations of motion w.r.t. $\vp^i$ resulting from \rf{string-action} :
\be
\vareps^{ab} \pa_{b} \vp^{i} \pa_{a} \Bigl(
\g^{cd} \pa_{c} X^{\m}\pa_{d} X^{\n} G_{\m\n}(X) -
\frac {\vareps^{cd}}{\sqrt{-\g}} F_{cd} \Bigr) = 0
\lab{vp-eqs}
\ee
imply (provided $\P (\vp) \neq 0$) :
\be
\g^{cd} \pa_{c} X^{\m}\pa_{d} X^{\n} G_{\m\n}(X) -
\frac {\vareps^{cd}}{\sqrt{-\g}} F_{cd} = M \; \Bigl( = \mathrm{const}\Bigr)
\; .
\lab{vp-eqs-a}
\ee
The equations of motion w.r.t. $\g^{ab}$ are:
\be
T_{ab} \equiv \pa_{a} X^{\m}\pa_{b} X^{\n} G_{\m\n}(X)   
- \h\g_{ab} \frac{\vareps^{cd}}{\sqrt{-\g}}F_{cd} = 0 \; .
\lab{gamma-eqs}
\ee
Both Eqs.\rf{vp-eqs-a}--\rf{gamma-eqs} yield $M=0$ and 
$ \Bigl( \pa_{a} X^{\m}\pa_{b} X^{\n} - 
\h \g_{ab} \g^{cd} \pa_{c} X^{\m}\pa_{d} X^{\n} \Bigr) G_{\m\n}(X) = 0$ ,
which is the same as in standard Polyakov-type formulation \rf{gamma-eqs-Pol}.

The equations of motion w.r.t. $X^\m$ read:
\be
\pa_a \(\P \g^{ab}\pa_b X^\m\) +
\P \g^{ab}\pa_a X^\n \pa_b X^\l \G^\m_{\n\l} = 0  \; ,
\lab{X-eqs}
\ee
where again $\G^\m_{\n\l}$ is the affine connection corresponding to the 
external space-time metric $G_{\m\n}$ as in the standard string case 
\rf{X-eqs-Pol}.

Most importantly, the equations of motion w.r.t. $A_a$ resulting from 
\rf{string-action} yield:
\be
\vareps^{ab} \pa_{b} \Bigl(\frac{\P (\vp)}{\sqrt{-\g}}\Bigr) = 0 \; ,
\lab{A-eqs}
\ee
which can be integrated to yield a {\em spontaneously induced} string tension: 
\br
\frac {\P (\vp)}{\sqrt{-\gamma}} = \textrm{const} \equiv T \; .
\nonu
\er

Since the modified-measure string model \rf{string-action} naturally
requires the presence of the auxiliary Abelian world-sheet gauge field $A_a$,
we may extend it by introducing a coupling of $A_a$ to some world-sheet
charge current $j^a$ :
\be
S = - \int d^2\s \,\P (\vp) \Bigl\lb \h \g^{ab} \pa_a X^{\m} \pa_b X^{\n} G_{\m\n}
- \frac{\vareps^{ab}}{2\sqrt{-\g}} F_{ab}(A)\Bigr\rb 
+ \int A_a j^a \; .
\lab{string-action-plus}
\ee
In particular, we may take $j^a$ to be the current of point-like charges on
the string, so that in the ``static'' gauge:
\be
\int A_a j^a =  - \sum_i e_{i} \int d\t A_0 (\t,\s_{i}) \; ,
\lab{static-gauge-term}
\ee
where $\s_i$ ($0 < \s_1 <\ldots < \s_N \leq 2\pi$) are the locations of the
charges. Now, the action \rf{string-action-plus} produces $A_a$-equations of
motion:
\be
\vareps^{ab}\pa_b E + j^a = 0 \quad ,\quad E \equiv \frac{\P (\vp)}{\sqrt{-\g}}
\; .
\lab{A-eqs-a}
\ee
Eqs.\rf{A-eqs-a} look exactly as $D=2$ Maxwell equations where the {\em variable} 
dynamical string tension $E\equiv \P (\vp)/\sqrt{-\gamma}$ is identified as 
world-sheet electric field strength, \textsl{i.e.}, canonically conjugated
momentum w.r.t. $A_1$ (the latter fact can be directly verified from the
explicit form of the $A$-term in the action \rf{string-action} or
\rf{string-action-plus}). 
\lskip
\textbf{Remark on Canonical Hamiltonian Treatment}. 
Introducing the canonical momenta resulting from the action
\rf{string-action-plus} :
\br
\pi^{\vp}_i = - \vareps_{ij} \pa_\s \vp^j 
\Bigl\lb \h \g^{ab} \pa_a X^{\m} \pa_b X^{\n} G_{\m\n}
- \frac{\vareps^{ab}}{2\sqrt{-\g}} F_{ab}(A)\Bigr\rb \; ,
\lab{can-mom-a} \\
\pi_{A_1} \equiv E = \frac{\P (\vp)}{\sqrt{-\g}} \;\; ,\;\;
\cP_\m = - \P (\vp) \(\g^{00}\dot{X}^\n + \g^{01}\pa_\s X^\n\) G_{\m\n} \; ,
\lab{can-mom-b}
\er
we obtain the canonical Hamiltonian as a linear combination of first-class 
constraints only. Part of the latter resemble the constraints in the ordinary
string case $\pi_{\g^{ab}}=0 $ and
\br
\cT_{\pm} \equiv \frac{1}{4} G^{\m\n}
\Bigl(\frac{\cP_\m}{E} \pm G_{\m\k} \pa_\s X^\k \Bigr)
\Bigl(\frac{\cP_\n}{E} \pm G_{\n\l} \pa_\s X^\l \Bigr) = 0 \; ,
\nonu
\er
where in the last Virasoro constraints the dynamical string tension $E$
appears instead of the {\em ad hoc} constant tension.

The rest of the Hamiltonian constraints are $\pi_{A_0} = 0$ and
\be
\pa_\s E - \sum_i e_{i} \d (\s - \s_{i}) = 0 \; ,
\lab{Gauss-law}
\ee
\textsl{i.e.}, the $D=2$ ``Gauss law'' constraint for the dynamical string
tension, which coincides with the $0$-th component of Eq.\rf{A-eqs-a}.
Finally, we have constraints involving only the measure-density fields:
\be
\pa_\s \vp^i \pi^{\vp}_i = 0 \quad ,\quad
\frac{\pi^{\vp}_2}{\pa_\s \vp^1} = 0  \; .
\lab{vp-constr}
\ee
The last two constraints span a closed Poisson-bracket algebra:
\br
\Pbr{\pa_\s \vp^i \pi^{\vp}_i (\s)}{\pa_{\s^\pr} \vp^i \pi^{\vp}_i (\s^\pr)} =
2 \pa_\s \vp^i \pi^{\vp}_i (\s) \pa_\s \d (\s -\s^\pr) + 
\pa_\s \(\pa_\s \vp^i \pi^{\vp}_i\) \d (\s -\s^\pr) \; ,
\nonu
\er
(a centerless Virasoro algebra), and:
\br
\Pbr{\pa_\s \vp^i \pi^{\vp}_i (\s)}{\frac{\pi^{\vp}_2}{\pa_\s \vp^1}(\s^\pr)} =
- \pa_\s \Bigl(\frac{\pi^{\vp}_2}{\pa_\s \vp^1}\Bigr) \d (\s -\s^\pr) \; .
\nonu
\er
Therefore, the constraints \rf{vp-constr} imply that the measure-density 
scalars $\vp^i$ are pure-gauge degrees of freedom.

\section{Classical Confinement Mechanism of ``Color'' Charges via Dynamical 
String Tension}

\subsection{Non-Abelian Generalization}

First, let us notice the following identity in $D=2$ involving Abelian gauge 
field $A_a$:
\be
{1\o {2\sqrt{-\g}}}\vareps^{ab} F_{ab}(A) = 
\sqrt{\h F_{ab}(A) F_{cd}(A) \g^{ac}\g^{bd}}  \; .
\lab{D2-ident}
\ee
This suggests the proper extension of the modified-measure bosonic string
model \rf{string-action} by introducing a {\em non-Abelian} (\textsl{e.g.},
$SU(\cN)$) auxiliary gauge field $A_a$ (here we take for simplicity flat 
external metric $G_{\m\n}=\eta_{\m\n}$) :
\br
S = - \int d^2\s \,\P (\vp) \Bigl\lb \h \g^{ab} \pa_a X^{\m} \pa_b X_\m 
-\sqrt{\h \Tr (F_{ab}(A)F_{cd}(A)) \g^{ac}\g^{bd}}\Bigr\rb
\nonu  \\
= - \int d^2\s \,\P (\vp) \Bigl\lb \h \g^{ab} \pa_a X^{\m} \pa_b X_\m 
- \frac{1}{\sqrt{-\g}} \sqrt{\Tr (F_{01}(A)F_{01}(A))}\Bigr\rb \; ,
\lab{string-action-NA}
\er
where $F_{ab}(A) = \pa_a A_b - \pa_b A_c + i \bigl\lb A_a,\, A_b\bigr\rb$.

The action \rf{string-action-NA} is again invariant under the {\em $\P$-extended Weyl
(conformal)} symmetry \rf{Phi-Weyl-symm}.

Notice that the {\em ``square-root'' Yang-Mills} action (with the regular
Riemannian-metric integration measure):
\be
\int d^2\s\, \sqrt{-\g} \sqrt{\h \Tr (F_{ab}(A) F_{cd}(A)) \g^{ac}\g^{bd}}
= \int d^2\s\, \sqrt{\Tr (F_{01}(A) F_{01}(A))}
\lab{sqrt-YM}
\ee
is a ``topological'' action similarly to the $D=3$ Chern-Simmons action
(\textsl{i.e.}, it is metric-independent).

Similarly to the Abelian case \rf{string-action-plus} we can also add a coupling of 
the auxiliary non-Abelian gauge field $A_a$ to an external ``color''-charge
world-sheet current $j^a$:
\be
S = - \int d^2\s \,\P (\vp) \Bigl\lb \h \g^{ab} \pa_a X^{\m} \pa_b X_\m -
- \frac{1}{\sqrt{-\g}} \sqrt{\Tr (F_{01}(A)F_{01}(A))}\Bigr\rb
+ \int \Tr\( A_a j^a\) \; .
\lab{string-action-NA-plus}
\ee
In particular, for a current of ``color'' point-like charges on the
world-sheet in the ``static'' gauge :
\be
\int \Tr\( A_a j^a\) = - \sum_{i}\Tr C_{i}\int d\t A_{0}(\t ,\s_{i}) \; ,
\lab{static-gauge-term-NA}
\ee
where $\s_i$ ($0 < \s_1 <\ldots < \s_N \leq 2\pi$) are the locations of the
charges.

The action \rf{string-action-NA-plus} produces the following equations of 
motion w.r.t. $\vp^i$ and $\g^{ab}$, respectively:
\be
\h \g^{cd}\pa_c X^\m \pa_d X_\m -\frac{1}{\sqrt{-\g}}
\sqrt{\Tr (F_{01} F_{01})} = M \; \Bigl( = \mathrm{const}\Bigr) \; ,
\lab{vp-eqs-NA}
\ee
\be
T_{ab} \equiv 
\pa_a X^\m \pa_b X_\m -\frac{1}{\sqrt{-\g}}\g_{ab}\sqrt{\Tr (F_{01} F_{01})}=0
\; .
\lab{gamma-eqs-NA}
\ee
As in the Abelian case the above Eqs.\rf{vp-eqs-NA}--\rf{gamma-eqs-NA} imply 
$M=0$ and the Polyakov-type equation \rf{gamma-eqs-Pol}.

The equations of motion w.r.t. auxiliary gauge field $A_a$ resulting from
\rf{string-action-NA-plus} resemble, similarly to the Abelian case \rf{A-eqs-a},
the $D=2$ non-Abelian Yang-Mills equations:
\be
\vareps^{ab} \nabla_a \cE + j^a = 0  \; ,
\lab{A-eqs-NA}
\ee
where:
\be
\nabla_a \cE \equiv \pa_a \cE + i \bigl\lb A_a,\,\cE \bigr\rb  \quad ,\quad
\cE \equiv \pi_{A_1} \equiv \frac{\P (\vp)}{\sqrt{-\g}}
\frac{F_{01}}{\sqrt{\Tr (F_{01} F_{01})}} \; .
\lab{E-def-NA}
\ee
Here $\cE$ is the non-Abelian electric field-strength -- the canonically 
conjugated momentum $\pi_{A_1}$ of $A_1$, whose norm is the dynamical string
tension $T \equiv |\cE| = \P (\vp)/\sqrt{-\g}$.

The equations of motion for the dynamical string tension following from 
\rf{A-eqs-NA} is:
\be
\pa_a \Bigl(\frac{\P (\vp)}{\sqrt{-\g}}\Bigr) + 
\vareps_{ab} \frac{\Tr\( F_{01} j^b\)}{\sqrt{\Tr\( F_{01}^2\)}} = 0 \; .
\lab{Tension-eqs}
\ee
In particular, the absence of external charges ($j^a =0$) :
$T \equiv \P (\vp)/\sqrt{-\g} = T_0 \equiv \mathrm{const}$

Finally, the $X^\m$-equations of motion $\pa_a \(\P (\vp) \g^{ab} \pa_b X^\m\)=0$
resulting from the action \rf{string-action-NA-plus} can be rewritten in the 
conformal gauge $\sqrt{-\g} \g^{ab} = \eta^{ab}$ as:
\be
\frac{\P (\vp)}{\sqrt{-\g}} \pa^a \pa_a X^\m  - {\wti j}^a \pa_a X^\m = 0
\quad ,\;\; \mathrm{where} \;\;
{\wti j}^a \equiv \frac{\Tr\( F_{01} j^a\)}{\sqrt{\Tr\( F_{01}^2\)}} \; .
\lab{X-eqs-NA}
\ee

For static charges ${\wti j}^0 = - \sum_i {\wti e}_i \d (\s -\s_i)$ :
\be
T \equiv \P (\vp)/\sqrt{-\g} = T_0 + \sum_i {\wti e}_i \th (\s -\s_i) 
\quad ,\quad
{\wti e}_i \equiv \frac{\Tr\( F_{01} C_i\)}{\sqrt{\Tr\( F_{01}^2\)}}\Bgv_{\s =\s_i}
\; ;
\lab{Tension-sol}
\ee
\be
T \pa^a \pa_a X^\m + \Bigl(\sum_i {\wti e}_i \d (\s -\s_i)\Bigr)\pa_\s X^\m = 0
\;\; \to
\left\{\begin{array}{ll}
\pa^a \pa_a X^\m = 0 \\
\pa_\s X^\m \Bgv_{\s =\s_i} = 0
\end{array} \right. \; .
\lab{X-sol}
\ee

\subsection{Classical Confinement Mechanism}

Recall that the modified string action \rf{string-action-NA-plus} yields
the $D=2$ Yang-Mills-like Eqs.\rf{A-eqs-NA} whose $0$-th component
$\pa_\s \cE + i\Sbr{A_1}{\cE} + j^0 = 0$ is the ``Gauss law'' constraint
for the dynamical string tension ($T \equiv |\cE| = \P (\vp)/\sqrt{-\g}$).
For point-like ``color'' charges and taking the gauge $A_1 =0$ (\textsl{i.e.},
$\cE \to {\wti \cE} = G \cE G^{-1}$ where $A_1 = -i G^{-1} \pa_\s G$),
the latter reads:
\be
\pa_\s {\wti \cE} -\sum_{i} {\wti C}_{i} \d (\s - \s_{i}) = 0 \quad ,\quad
{\wti C}_{i} \equiv G C_i G^{-1} \Bgv_{\s =\s_i} \; .
\lab{Gauss-law-NA}
\ee

Let us consider the case of {\em closed} modified string with positions of the
``color'' charges at $0 <\s_1 <\ldots <\s_N \leq 2\pi$. Then, integrating the
``Gauss law'' constraint \rf{Gauss-law-NA} along the string (at fixed proper
time) we obtain:
\be
\sum_i {\wti C}_i = 0 \quad ,\quad {\wti \cE}_{i,i+1} = {\wti \cE}_{i-1,i} + 
{\wti C}_i \; ,
\lab{charge-constr-NA}
\ee
where ${\wti \cE}_{i,i+1} = {\wti \cE}$ in the interval $\s_i < \s <\s_{i+1}$. 

The discussion in this section leads to the following conclusions:
\begin{itemize}
\item
We see from Eqs.\rf{Tension-sol}--\rf{X-sol} that the modified-measure
(closed) string with $N$ point-like (``color'') charges on it 
(\rf{string-action-plus} or \rf{string-action-NA-plus}) is equivalent to $N$ 
chain-wise connected regular open string segments obeying Neumann boundary
conditions. 
\item
Each of the above open string segments, with end-points at the charges
$e_{i}$ and $e_{i+1}$ (in the Abelian case) or $C_{i}$ and $C_{i+1}$ (in the 
non-Abelian case), has {\em different} constant string tension $T_{i,i+1}$
such that $T_{i,i+1} = T_{i-1,i} + \stackrel{(\sim )}{e}_i$ (the non-Abelian
${\wti e}_i$ are defined in \rf{Tension-sol}).
\item
Eq.\rf{charge-constr-NA} tells us that the only (classically) admissable 
configuration of ``color'' point-like charges coupled to a modified-measure 
closed bosonic string is the one with {\em zero} total ``color'' charge, 
\textsl{i.e.}, the model \rf{string-action-NA-plus} provides a classical 
mechanism of ``color'' charge confinement.
\end{itemize}

\section{Branes with a Modified World-Volume Integration Measure}

Before generalizing our construction from the previous two sections to the case
of higher-dimensional $p$-branes, let us recall the standard Polyakov-type 
formulation of the bosonic $p$-brane action:
\be
S = -{T\o 2}\int d^{p+1}\s\,\sqrt{-\g}\, \Bigl\lb 
\g^{ab}\pa_a X^\m \pa_b X^\n G_{\m\n}(X) - \L (p-1)\Bigr\rb \; .
\lab{stand-brane-action}
\ee
Here $\g_{ab}$ is the ordinary Riemannian metric on the $p+1$-dimensional
brane world-volume with $\g \equiv \det\v\v \g_{ab}\v\v$. The world-volume indices
$a,b=0,1,\ldots ,p$; $T$ is the given \textsl{ad hoc} brane tension; the constant
$\L$ can be absorbed by rescaling $T$ (see below Eq.\rf{stand-brane-action-NG}.
The equations of motion w.r.t. $\g^{ab}$ and $X^\m$ read:
\be
T_{ab} \equiv \( \pa_a X^\m \pa_b X^\n - 
\h \g_{ab} \g^{cd}\pa_c X^\m \pa_d X^\n \) G_{\m\n} + \g_{ab} {\L\o 2}(p-1)
= 0 \; ,
\lab{stand-brane-gamma-eqs}
\ee
\be
\pa_a \(\sqrt{-\g}\g^{ab}\pa_b X^\m\) +
\sqrt{-\g}\g^{ab}\pa_a X^\n \pa_b X^\l \G^\m_{\n\l} = 0  \; .
\lab{stand-brane-X-eqs}
\ee
Eqs.\rf{stand-brane-gamma-eqs} when $p \neq 1$ imply:
\be
\L \g_{ab} = \pa_a X^\m \pa_b X^\n G_{\m\n} \; ,
\lab{stand-brane-metric-eqs}
\ee
which in turn allows to rewrite Eq.\rf{stand-brane-gamma-eqs} as:
\be
T_{ab} \equiv \( \pa_a X^\m \pa_b X^\n - 
{1\o {p+1}} \g_{ab} \g^{cd}\pa_c X^\m \pa_d X^\n \) G_{\m\n} = 0 \; .
\lab{stand-brane-gamma-eqs-Pol}
\ee
Furthermore, using \rf{stand-brane-metric-eqs} the Polyakov-type brane action 
\rf{stand-brane-action} becomes on-shell equivalent to the Nambu-Goto-type brane
action:
\be
S = - T \L^{-{{p-1}\o 2}} \int d^{p+1}\s\,
\sqrt{-\det\v\v \pa_a X^\m \pa_b X^\n G_{\m\n} \v\v} \; .
\lab{stand-brane-action-NG}
\ee

\subsection{Modified-Measure Brane Actions}

Now, similarly to the string case we introduce a modified world-volume
integration measure in terms of $p+1$ auxiliary scalar fields $\vp^i$
($i=1,\ldots ,p+1$) :
\be
\P (\vp) \equiv \frac{1}{(p+1)!} \vareps_{i_1\ldots i_{p+1}} 
\vareps^{a_1\ldots a_{p+1}} \pa_{a_1} \vp^{i_1}\ldots \pa_{a_{p+1}} 
\vp^{i_{p+1}} \; ,
\lab{brane-measure}
\ee
and consider the following modified $p$-brane action:
\be
S = - \int d^{p+1}\s\, \P (\vp) \Bigl\lb \h \g^{ab} \pa_a X^\m \pa_b X^\n G_{\m\n}(X)
+ {1\o \sqrt{-\g}}\O (A) \Bigr\rb + \int d^{p+1}\s\,\cL (A)  \; .
\lab{brane-action}
\ee
The term $\O (A)$ indicates a topological density given in terms of some auxiliary 
gauge (or matter) fields $A^I$ living on the world-volume, 
``topological'' meaning that:
\be
\partder{\O}{A^I} - \pa_a \(\partder{\O}{\pa_a A^I}\) = 0 \;\;
\mathrm{identically} \quad ,\quad
\mathrm{i.e.}\;\; \d\O (A) = \pa_a \(\partder{\O}{\pa_a A^I} \d A^I\) \; . 
\lab{top-density-def}
\ee
$\cL (A)$ describes possible coupling of the auxiliary fields $A^I$ to
external ``currents'' on the brane world-volume.

The requirement for $\O (A)$ to be a topological density is dictated by the
requirement that the modified-measure brane action \rf{brane-action} (in the 
absence of the last gauge/matter term $\int d^{p+1}\s\,\cL (A)$) reproduces 
the ordinary $p$-brane equations of motion apart from the fact that the brane 
tension $T \equiv \P (\vp)/\sqrt{-\g}$ is now
an {\em additional dynamical degree of freedom}.

The simplest example of a topological density $\O (A)$ for the auxiliary
gauge/matter fields is:
\be
\O (A) = -\frac{\vareps^{a_1\ldots a_{p+1}}}{p+1} F_{a_1\ldots a_{p+1}} (A)
\quad ,\quad
F_{a_1\ldots a_{p+1}} (A) = (p+1)\pa_{\lb a_1} A_{a_2\ldots a_{p+1}\rb} \; ,
\lab{top-density-p-rank}
\ee
where $A_{a_1 \ldots a_p}$ is rank $p$~ antisymmetric tensor (Abelian) gauge
field on the world-volume\foot{A modified $p$-brane model significantly
different from \rf{brane-action}--\rf{top-density-p-rank} has been proposed in 
ref.\ct{Bergshoeff-etal}. The latter model also contains world-volume $p$-form 
gauge fields which, however, appear quadratically in the brane action of
\ct{Bergshoeff-etal} and, therefore, they are dynamical rather than auxiliary
fields.}.

More generally, for $p+1=rs$ we can have:
\be
\O (A) = {1\o {rs}}\vareps^{a_{11}\ldots a_{1r} \ldots a_{s1}\ldots a_{sr}} 
F_{a_{11}\ldots a_{1r}} \ldots F_{a_{s1}\ldots a_{sr}} \; .
\lab{top-density-rs}
\ee

We may also employ {\em non-Abelian} auxiliary gauge fields as in the string case.
For instance, when $p=3$ we may take:
\be
\O (A) = {1\o 4}\vareps^{abcd} \Tr\( F_{ab}(A) F_{cd}(A)\)
\lab{top-density-NA}
\ee
or, more generally, for $p+1=2q$ :
\be
\O (A) = 
{1\o {2q}}\vareps^{a_1 b_1\ldots a_q b_q} \Tr\( F_{a_1 b_1}\ldots F_{a_q b_q}\)
\; ,
\lab{top-density-NA-q}
\ee
where $F_{ab}(A) = \pa_a A_b - \pa_b A_c + i \bigl\lb A_a,\, A_b\bigr\rb$.

The modified $p$-brane action \rf{brane-action} produces the following equations
of motion w.r.t. $\vp^i$ :
\be
\h \g^{cd} \pa_c X^\m \pa_d X^\n G_{\m\n} + {1\o \sqrt{-\g}}\O (A) 
= M \equiv \mathrm{const} \; ,
\lab{brane-vp-eqs}
\ee
and w.r.t. $\g^{ab}$ (assuming that $\int d^{p+1}\s\,\cL (A)$ does
not depend on $\g_{ab}$ -- true \textsl{e.g.} if it describes coupling of the
auxiliary (gauge) fields $A$ to charged lower-dimensional branes) :
\be
\pa_a X^\m \pa_b X^\n G_{\m\n} + \frac{\g_{ab}}{\sqrt{-\g}} \O (A) = 0 \; .
\lab{brane-gamma-eqs}
\ee
Both Eqs.\rf{brane-vp-eqs}--\rf{brane-gamma-eqs} imply:
\be
\O (A) = - \frac{2M}{p-1} \sqrt{-\g} \; ,
\lab{top-density-eq}
\ee
\be
\pa_a X^\m \pa_b X^\n G_{\m\n} = \g_{ab}\,\frac{2M}{p-1} \;\;\ ,\;\;
\pa_a X^\m \pa_b X^\n G_{\m\n}
- {1\o {p+1}} \g_{ab} \g^{cd} \pa_c X^\m \pa_d X^\n G_{\m\n} = 0  \; .
\lab{brane-eqs-Pol}
\ee
The last two Eqs.\rf{brane-eqs-Pol} reproduce two of the ordinary brane 
equations of motion \rf{stand-brane-metric-eqs}--\rf{stand-brane-gamma-eqs-Pol}
in the standard Polyakov-type formulation.

We now consider the modified brane \rf{brane-action} equations of motion w.r.t.
auxiliary (gauge) fields $A^I$ -- these are the eqs. determining the dynamical
brane tension $T \equiv \P (\vp)/\sqrt{-\g}$ :
\be
\pa_a \Bigl(\frac{\P (\vp)}{\sqrt{-\g}}\Bigr)\,\partder{\O}{\pa_a A^I} + j_I = 0
\; ,
\lab{brane-A-eqs}
\ee
where $j_I \equiv \partder{\cL}{A^I} - \pa_a \Bigl(\partder{\cL}{\pa_a A^I}\Bigr)$
is the corresponding ``current'' coupled to $A^I$.

As a physically interesting example let us take the choice \rf{top-density-p-rank}
for the topological density $\O (A)$ and consider the following natural coupling
of the auxiliary $p$-form gauge field:
\be
\int d^{p+1}\s\,\cL (A) = \int d^{p+1}\s\, A_{a_1\ldots a_p} j^{a_1\ldots a_p}
\lab{brane-A-coupling}
\ee
to an external world-volume current:
\be
j^{a_1\ldots a_p} = \sum_i e_i 
\int_{\cB_i} d^p u\, \frac{1}{p!} \vareps^{\a_1\ldots\a_p}
\partder{\s_i^{a_1}}{u^{\a_1}}\ldots \partder{\s_i^{a_p}}{u^{\a_p}}
\d^{(p+1)} \bigl(\underline{\s} - \underline{\s}_i (\underline{u})\bigr) \; .
\lab{p-form-current}
\ee
Here $j^{a_1\ldots a_p}$ is a current of charged $(p-1)$-sub-branes $\cB_i$ 
embedded into the original $p$-brane world-volume via
$\s^a = \s^a_i (\underline{u})$ with parameters
$\underline{u} \equiv (u^\a)_{\a =0,\ldots, p-1}$. For simplicity we assume
that the $\cB_i$ sub-branes do not intersect each other. With the choice
\rf{brane-A-coupling}--\rf{p-form-current}, Eq.\rf{brane-A-eqs} 
for the dynamical brane tension $T \equiv \P (\vp)/\sqrt{-\g}$ becomes:
\be
\pa_a \Bigl(\frac{\P (\vp)}{\sqrt{-\g}}\Bigr) + \sum_i e_i \cN^{(i)}_a = 0
\; ,
\lab{brane-tension-eq}
\ee
where $\cN^{(i)}_a$ is the normal vector w.r.t. world-hypersurface of the 
$(p-1)$-sub-brane $\cB_i$ :
\be
\cN^{(i)}_a \equiv {1\o {p!}} \vareps_{a b_1\ldots b_p}
\int_{\cB_i} d^p u\, \frac{1}{p!} \vareps^{\a_1\ldots\a_p}
\partder{\s_i^{a_1}}{u^{\a_1}}\ldots \partder{\s_i^{a_p}}{u^{\a_p}}
\d^{(p+1)} \bigl(\underline{\s} - \underline{\s}_i (\underline{u})\bigr)
\; .
\lab{def-N}
\ee

Finally, the modified-brane action \rf{brane-action} yields the
$X^\m$-equations of motion $\pa_a \(\P (\vp) \g^{ab} \pa_b X^\m \) = 0$
(taking for simplicity $G_{\m\n} = \eta_{\m\n}$), which upon using 
\rf{brane-tension-eq} can be rewritten in the form
\foot{For a detailed description of techniques for obtaining solutions of
equations of motion for standard string and brane systems in non-trivial
backgrounds, which can be easily adapted in the present modified-measure 
string and brane models, see ref.\ct{Plamen}.} :
\be
\frac{\P (\vp)}{\sqrt{-\g}}\pa_a \(\sqrt{-\g} \g^{ab}\pa_b X^\m\) 
-\sum_i e_i \cN^{(i)}_a \g^{ab}\pa_b X^\m = 0  \; .
\lab{brane-X-eqs}
\ee

\subsection{Confinement of Charged Lower-Dimensional Branes}

Let us consider the solutions for the for the dynamical brane tension
Eq.\rf{brane-tension-eq}. Recalling the definition \rf{def-N} of $\cN^{(i)}_a$
we find from \rf{brane-tension-eq} that $T\equiv \P (\vp)/\sqrt{-\g}$ is 
piece-wise constant on the $p$-brane world-volume with jumps when crossing the 
world-hypersurface of each charged $(p-1)$-sub-brane $\cB_i$, the
corresponding jump being equal to the charge magnitude $\pm e_i$ (the
overall sign depending on the direction of crossing w.r.t. the normal
$\cN^{(i)}_a$).

Taking into account the above piece-wise constant solution for
$T\equiv \P (\vp)/\sqrt{-\g}$, the $X^\m$-equations of motion 
\rf{brane-X-eqs} for the closed modified brane \rf{brane-action} become equivalent 
to the following set of equations of motion: 
\be
\pa_a \(\sqrt{-\g} \g^{ab}\pa_b X^\m\) = 0 \quad ,\quad 
\pa_N X^\m \Bgv_{\cB_i} = 0 \; ,
\lab{brane-X-eqs-segments}
\ee
where $\pa_N$ indicates normal derivative w.r.t. world-hypersurface of the 
$(p-1)$-sub-brane $\cB_i$. Therefore, Eqs.\rf{brane-X-eqs-segments}
together with Eq.\rf{brane-tension-eq} describe a set of 
{\em ordinary open} $p$-brane segments with common boundaries, where
each open $p$-brane segment possesses {\em different constant} brane tension
and obeys Neumann boundary conditions.

Integrating Eq.\rf{brane-tension-eq} along arbitrary smooth closed curve $\cC$
on the $p$-brane world-volume which is transversal to (some or all of) the
$(p-1)$-sub-brane $\cB_i$, we obtain the following constraints on the
possible sub-brane configurations:
\be
\sum_i e_i\, n_i (\cC) = 0 \; ,
\lab{charge-constr}
\ee
where $n_i (\cC)$ is the sign-weighted total number of $\cC$ crossing $\cB_i$.
In the present $p\geq 2$-brane case, however, due to the much more complicated
topologies of the pertinent world-volumes Eq.\rf{charge-constr} may yield various
different types of allowed sub-brane configurations. 

As a simple illustration, here we will only consider the simplest non-trivial case 
$p=2$ and take the static gauge for the $p=1$ sub-branes (strings), \textsl{i.e.},
the proper times of the charged strings coinsides with the proper time of the
bulk membrane. The latter means that the fixed-time world-volume of the bulk
{\em closed} membrane is a Riemann surface with some number $g$ of handles
and no holes. Further, we will assume the following simple topology of the
attached $N$ charged strings $\cB_i$ : upon cutting the membrane surface
along these attached strings it splits into $N$ {\em open} membranes $\cM_i$
($i=1,\ldots ,N$) with Neumann boundary conditions (cf. \rf{brane-X-eqs-segments}), 
each of which being a Riemann surface with $g_i$ handles and
2 holes (boundaries) formed by the strings $\cB_{i-1}$ and $\cB_i$,
respectively\foot{The Euler charasteristics of the bulk membrane Riemann
surface is $\chi = 2 -2g$, whereas for the open brane $\cM_i$ it is 
$\chi_i = 2 -2g_i -2$, so that $\chi = \sum_i \chi_i$ or, equivalently,
$g = 1 + \sum_i g_i$.}. The brane tension of $\cM_i$ is a dynamically
generated constant $T_i$ where $T_{i+1} = T_i + e_i$. In the present 
configuration Eq.\rf{charge-constr} evidently reduces to the
constraint $\sum_i e_i = 0$.

Thus, we conclude that similarly to the string case, modified-measure
$p$-brane models describe configurations of charged $(p-1)$-branes with
charge confinement. Apart from the latter, in general there exist more
complicated configurations allowed by the constraint \rf{charge-constr},
which will be studied elsewhere.


\section{Conclusions}

The above discussion shows that there exist natural from physical point of
view modifications of world-sheet and world-volume integration measures which
may significantly affect string and brane dynamics. Let us sumarize the main
features of the new modified-measure string and brane models:
\begin{itemize}
\item
Acceptable dynamics {\em naturally} requires the introduction of auxiliary
world-sheet gauge field (world-volume $p$-form tensor gauge field).
\item
The string/brane tension is {\em not} anymore a constant scale given {\em ad hoc},
but rather appears as an {\em additional dynamical degree of freedom} beyond the
ordinary string/brane degrees of freedom.
\item
The dynamical string/brane tension has physical meaning of an electric field
strenght for the auxiliary gauge field.
\item
The dynamical string/brane tension obeys ``Gauss law'' constraint equation
and may be nontrivially variable in the presence of point-like charges (on
the string world-sheet) or charged lower-dimensional branes (on the
$p$-brane world-volume).
\item
Modified-measure string/brane models provide simple classical mechanisms for
confinement of point-like ``color'' charges or charged lower-dimensional branes
due to variable dynamical tension.
\end{itemize}


{\bf Acknowledgement.} E.N. and S.P. are partially 
supported by Bulgarian NSF grant \textsl{F-904/99}.



\bigskip

\begin{thebibliography}{9}
\small
\bi{string-brane-rev}
Ne'eman, Y. and Eizenberg, E. (1995), {\em ``Membranes and Other Extendons''}, 
World Scientific; 
Green, M., Schwarz, J. and Witten, E. (1987), {\em ``Superstring Theory''}, 
Vol.1,2, Cambridge Univ. Press; 
Polchinksi, J. (1998), {\em ``String Theory''}, Vol.1,2, Cambridge Univ. Press.
\bi{TMT}
Guendelman E.I. and Kaganovich, A. \PRD{53}{1996}{7020};
Guendelman E.I. and Kaganovich, A. \PRD{60}{1999}{065004};
Guendelman E.I. (2000), \textsl{Class. Quant. Grav.} \textbf{17} 261;
Guendelman E.I. (2001), \textsl{Found. Phys.} \textbf{31} 1019;
Kaganovich, A. \PRD{63}{2001}{025022};
Guendelman E.I. and Kaganovich, A. (2002), \textsl{Int. J. Mod. Phys.}
\textbf{A17} 417 ~(\textsl{hep-th/0106152});
Guendelman E.I. and Kaganovich, A. (2002),
\textsl{Mod. Phys. Lett.} \textbf{A17} 1227 ~(hep-th/0110221).
\bi{TMT-Kiten}
Guendelman E.I. and Kaganovich, A. (2002), this volume.
\bi{mstring}
Guendelman E.I. \CQG{17}{2000}{3673} ~(hep-th/0005041);
Guendelman E.I. \PRD{63}{2001}{046006} ~(hep-th/0006079);
Guendelman E.I., Kaganovich, A., Nissimov, E. and Pacheva, S. (2002),
\textsl{Phys. Rev.} \textbf{D66} 046003 ~(hep-th/0203024).
\bi{Polyakov}
Polyakov, A. \PLB{103}{1981}{207,211}.
\bi{K-V}
Katanaev, M. and Volovich, I. (1986), \textsl{Phys. Lett.} \textbf{B175} 413
~(hep-th/0209014).
\bi{Bergshoeff-etal}
Bergshoeff, E., London, L. and Townsend, P.K. \CQG{9}{1992}{2545} 
~(hep-th/9206026).
\bi{Plamen}
Bozhilov, P. (2002), this volume; Bozhilov, P. (2001), hep-th/0111103; 
Bozhilov, P. \PRD{65}{2001}{026004} ~(hep-th/0103154); and references therein.

%
%
%
%

\end{thebibliography}
\end{document}